\documentclass[a4paper]{aa}

\usepackage{graphicx}
\usepackage{natbib}
\bibpunct{(}{)}{;}{a}{}{,}

\usepackage{txfonts}

\newcommand{\mr}[1]{\mathrm{#1}}

\title{Radial distribution of planets}

\subtitle{ Predictions based on the core-accretion gas-capture
planet-formation model}
\author{K. Kornet\inst{1,2} \and S. Wolf\inst{1}} 

\institute{Max Planck Institute for Astronomy, K\"onigstuhl 17, 69117
Heidelberg, Germany
\and
Nicolaus Copernicus Astronomical Center, Bartycka 18,
Warsaw, 00-716, Poland}

\abstract{We have investigated the problem of the distribution of both
masses and orbital radii of planets resulting from the gas-accretion,
gas-capture model.  First we followed the evolution of gas and solids
from the moment where all solids are in the form of small grains to
the stage when most of them are in the form of planetesimals for a set
of different initial masses and sizes of protoplanetary disks. Based
on that we performed Monte Carlo calculations describing the formation
of giant places at different locations. We included the effects of
type II migration and growth of the mass of the planet after the gap
opened. We discuss how these effects influence the final distribution
of giant planets. We show that when the giant planets are not able to
migrate or grow in mass after the gap opens, their
distribution is mainly determined by the properties of the gaseous
disk. However, with those two effects included, reproducing the
parameters of the gaseous disks from the distribution of planets
becomes difficult. We also checked the roles of both the material of which the
solids consist and the mass of the central star. The main result is
that, in disks around less massive stars, giant planets at the given
location tend to be less massive. At the same time, the giant planets
with the given mass tend to form closer to the less massive stars.}

\begin{document}

\maketitle

\section{Introduction}
Current radial velocity surveys have led to the discovery of over 150
extrasolar planets around main-sequence stars \citep[see][]{marcy05,mayor04}.
 The large majority of these planets  are probably gas giant
planets, as their masses are above $100 M_\oplus$. Such a collection
provides a good data set for comparing predictions of theories of planet
formation \citep[e.g.][]{ida04a,ida04b,alibert05,kac05}.

The standard model for the formation of giant planets is the
core-accretion, gas capture-model. The numerical calculations of
\citet{pollack96} show, that the formation of a giant planet in this
model can be divided into three phases. During the first one, the
solid core is formed by the collisional accumulation of
planetesimals. The second phase begins when the core reaches the mass
of a few Earth masses and starts to accrete a significant amount of
gas. During this phase the envelope stays in quasi - static and
thermal equilibrium, as the energy radiated by the envelope is
compensated for by the energy released by accreted planetesimals. As
during this phase the protoplanet accretes gas with a higher rate than
solids, the mass of the envelope finally becomes equal to the mass of
the core. At this moment phase 3 begins, during which the planet
rapidly grows in mass by runaway accretion of gas. The final mass and
location of a giant planet are determined by its gravitational
interaction with its environment. As it grows in mass it induces
spiral waves in the gaseous disk. This leads to the transfer of
angular momentum resulting in the inward migration of the planet and
possibly in the gap opening \citep{lin86,lin96,ward97}. This last
phenomenon strongly reduces the further growth of the planet.

The main problem with that scenario is related to the timescale
required to form a giant planet in it. In general it is the same order
of magnitude as the lifetime of the protoplanetary disk, and it is not
necessarily certain if the giant planet is able to form before
dispersion of the disk.  Close to the star, the formation time of a
giant planet in the gas capture model is determined by phase 2, while
at larger distances ($\gtrsim 10 \mr{AU}$) the lengths of phases 1 and
2 become comparable. The lengths of these two phases depend on the
initial surface density of the planetesimal swarm in the given
location. The larger the density, the faster the core grows and
reaches higher mass at the end of phase 1. With the higher mass of the
core, the length of phase 2 also diminishes. In general, at every
distance from the star there is a critical value of the surface
density of planetesimals which enables formation of a giant planet
before dispersion of the protoplanetary disk; for a more detailed
discussion see \citet{kac_mass}.

However, the density of the protoplanetary swarm is not in a simple
relation with the density of gas in the disk from which it
emerges. While the gaseous component evolves in a viscous way, the
evolution of the solid component is governed by processes of gas drag,
coagulation, sedimentation, and evaporation \citep{weiden93}.  A
significant redistribution of solid material takes place in the
result, and in the inner parts of the disk its surface density can be
significantly enhanced compared to the initial value
\citep{sv97,weiden03}. Consequently, analysis of the possible masses
and orbits of giant planets resulting from the core accretion scenario
should also include the global evolution of solids in protoplanetary
disks. A simple model of this evolution was proposed by \citet{kac1}
and further extended by \citet{kac2} and \citet{kac05} to include
subsequent formation of giant planets.

In this paper we extend our models to also include the effects of 
migration and the gap opening by the planet. It
enables us for the first time to characterize the distribution of
planetary masses and final locations resulting from the core accretion model,
not only including  the growth of the planet but also the preceding
evolution of solids, which determines the surface density of the
planetesimal swarm. In Sect. \ref{s:methods} we explain our approach
to the evolution of the protoplanetary disk and planet formation. The
results are presented in Sect. \ref{s:results} and discussed in Sect. \ref{s:concl}.

\section{Methods}
\label{s:methods}
In our approach we divide the formation of giant planets into three
phases in a natural way. In the first one the planetesimal swarm is
formed in the protoplanetary disk by the collisional accumulations of
solids. This phase lasts till the gravitational interactions between
solid bodies become the dominant factor governing their evolution. In
the second epoch the giant planet is build by the accretion of the
planetesimals, and at later stages of gas onto the protoplanetary
core. Finally, when the mass of the planet reaches a critical value,
the planet migrates within the disk and reaches its final position. We now
discuss how we modeled each of these phases.

\subsection{Formation of a planetesimal swarm}

We describe the protoplanetary disk as a two-component fluid
consisting of gas and solids. We modeled the gaseous component as a
geometrically thin turbulent $\alpha$ disk \citep{73aa24_337}. Its
surface density $\Sigma$ is given as a function of distance $a$ from
the star and time $t$ in terms of a selfsimilar solution of
\citet{s98}. All other quantities characterizing the gas were obtained
by solving the standard set of equations for a thin-disk approximation
\citep[e.g.][]{accr_power}. The initial conditions were parameterized
by two quantities: the initial mass of the disk $M_0$ and its initial
outer radius $R_0$.

The crucial approximation underlying our approach to the evolution of
solids is that the size distribution of particles at any given radial
location is narrowly peaked around a mean value particular to this
location and time. In the practical implementations it means that
sizes of the particles $s$ are expressed as a single-value function of
time and position, $s=s(t,r)$.
 Collectively, all solid particles are
treated as a turbulent fluid characterized by its mean surface
density $\Sigma_{\rm s}(t,r)$, which is governed by the continuity equation:
\begin{equation}
\frac{\partial \Sigma_{\rm s}}{\partial t} +
\frac{1}{r}\frac{\partial}{\partial r}(r V_{{\rm s}} \Sigma_{\rm s})=0
\mr{.}
\label{dustEvolution}
\end{equation}
The radial drift velocity of solids $V_{\rm s}$ is the result of the
frictional coupling between solids and the gas and depends on the local
size of the particles and properties of the gas. In the regions where
the temperature of the disk is higher than the evaporation temperature
$T_\mathrm{evap}$, the solids are treated as vapour with a velocity
equal to the velocity of the gas.

The size of the
particles is governed by the second equation:
\begin{equation}
\frac{\partial \Sigma_{\rm size}}{\partial t} +
\frac{1}{r}\frac{\partial}{\partial r}(r V_{{\rm s}} \Sigma_{\rm size}) = f \Sigma_{\rm s}
\label{sizeDustEvolution}
\end{equation}
where $\Sigma_\mathrm{size}=s \Sigma_{\rm s}$.  The source function $f$
describes the growth of particles due to mutual collisions and
subsequent coagulation. 
The main assumptions used in its derivation are that all collisions
between particles lead to coagulation and that the relative velocities
of colliding particles are given by the turbulent model described by
\citet{sv97}.  In the calculation of the density of solids from their surface
density, the effect of their sedimentation toward the midplane of the
disk is taken into account by evolving the scale height of the solid disk.

Initially the surface density of solids is everywhere a constant
fraction of the surface density of gas, and their sizes amount to
$10^{-3}\ \mathrm{cm}$. The results discussed in the subsequent
sections do not depend on the choice of that particular value, as long
as the solids are initially small enough to couple well to the gas.

\subsection{Growth of planets}
\label{s:growth}
Our procedure for the formation of the giant planet starts when the
planetesimals at a given point in the disk reach radii of $2\
\mr{km}$. That moment determines the initial surface density of the
planetesimal swarm.  The growth of the protoplanetary core is described by
the following formula given by \citet{papaloizou99}:
\begin{equation}
\dot{M_\mathrm{c}}=C_1 C_\mathrm{cap} R_\mathrm{p} R_\mathrm{H} \Omega_\mathrm{K}
\Sigma_\mathrm{s}
\label{eq:mc}
\end{equation}
where
\begin{equation}
R_\mathrm{H}=a\left(\frac{M_\mathrm{p}}{3 M_\star}\right)^{1/3}
\end{equation}
is the radius of the Hill sphere of the planet and $\Omega_\mr{K}$ the
Keplerian angular velocity.  The value of $C_1$ given by
\citet{papaloizou99} is $81\pi/32$; we use a factor of 5 (the
difference comes from the different definition of $R_\mathrm{H}$).
The quantity $C_\mathrm{cap}$ describes the increase in the effective
capture radius of the planet within respect to its real radius
$R_\mathrm{p}$ due to the interaction of planetesimals with the
envelope of the planet \citep{podolak88}. We use for it an approximate
fit provided by \citet{hubickyj01}, made to the results of
\citet{boden00}. For those core masses that are less then $5
M_{\oplus}$, no increase in the effective capture radius is assumed,
i.e. $C_\mathrm{cap}(M_\mathrm{c}<5M_\oplus)=1$. For larger core
masses, it increases linearly with the mass of the core, reaching its
maximum value of $C_\mathrm{cap}=5$ for $M_\mathrm{c}=15 M_\oplus$. We
assume that the surface density of planetesimals $\Sigma_\mr{d}$ is
constant throughout the feeding zone which, extends to 4 Hill radii on
both sides of the planetary orbit. Furthermore, $\Sigma_\mr{d}$
changes only due to accretion of the planetesimals onto the core and
the growth of the feeding zone.

Significant accretion of gas onto the core begins when it reaches
a critical mass of
\begin{equation}
M_\mathrm{c,crit}= 10 \left(\frac{\dot{M}_\mathrm{c}}{10^{-6}
  \ M_\oplus\ yr^{-1}}\right)^{0.25}
\left(\frac{\kappa_\mathrm{env}}{1\ \mathrm{cm}^2\ \mathrm{g}^{-1}}\right)^{0.25}
\label{e:mcrit}
\end{equation}
\citep{ikoma00}, where $\kappa_\mathrm{env}$ is the opacity in the
envelope of the planet. Its actual magnitude is currently poorly
constrained. We assume that $\kappa_\mathrm{env}=1\ \mathrm{cm}^2\
\mathrm{g}^{-1}$. When the mass of the protoplanet exceeds
$M_\mathrm{c,crit}$, it contracts on the Kelvin-Helmholtz time scale
$\tau_\mathrm{KH}$.  By fitting the result of
\citet{pollack96} \citet{bryden00} show that
\begin{equation}
\tau_\mathrm{KH}=10^{b} \left(\frac{M_\mathrm{p}}{M_\oplus}\right)^{-c}
    \left(\frac{\kappa}{1 \mathrm{cm^2\ g^{-1}}}\right)\ \mathrm{yr}
\end{equation}
where the exponents  $b=10$ and $c=3$. However, the gas accretion
rate connected with this contraction cannot be higher than the speed
with which viscous transport of gas replenishes the feeding zone of the
planet.  Consequently, we adopt the
following equations for the gas accretion rate onto the star
\begin{equation}
\frac{\mathrm{d} M_\mathrm{env}}{\mathrm{d}t} =
  \min \left[\frac{M_\mathrm{p}}{\tau_\mathrm{KH}}, \dot{M}_\mr{disk} \right]
\label{eq:menv}
\end{equation}
where $\dot{M}_\mr{disk}$ is the accretion rate in
the disk without the planet. We assume that it is equal to $\pi
\alpha C_s H \Sigma$, where $C_S$ is the sound speed in the gaseous
disk, $H$ its scale height, and $\Sigma$ its gas surface density at the given
location. 

When the Hill radius of the planet becomes larger than 1.5 of the scale
height of the disk, the planet induces a strong tidal torque on the
disk and opens a gap in it. As a  result the accretion of gas is
strongly diminished. Its maximum rate is then equal to:
\begin{equation}
\dot{M}_\mr{env}= \dot{M}_\mr{disk} \left\{1.668 
\left(\frac{M_\mr{p}}{M_\mr{J}}\right)^{1/3} \exp \left[-\frac{M_\mr{p}}{1.5
M_\mr{J}}\right]+0.04\right\}
\end{equation}
where $M_\mr{J}$ is the Jupiter mass \citep{veras04}.

\subsection{Migration}
\label{s:migr}
The gravitational interaction of a planet with the disk leads to its
migration and to formation of the gap in the disk \citep{lin86,ward97}.
For a low-mass planet the interaction is linear and results in 
so-called type~I migration. In contrast to this scenario, high-mass planets
open a gap in the disk that reduces the timescale of migration
(referred to as type II migration).

Analytical calculations show that the timescale of type I migration in
a laminar disk is much shorter than both the disk lifetime and the
timescale of planet formation. Consequently, if type I migration has taken
place, nearly all planets would be accreted onto the star and the
probability of finding any planetary system would be very
low. Likewise, recent simulations by \citet{nelson04} suggest that
the torques exerted onto a low mass planet in a turbulent MHD disk
fluctuate strongly. As a result the planet proceeds at a random walk
and the mean value of the migration velocity is highly reduced. Due to
these uncertainties we neglect the type I migration in our models.

Type II migration begins when the planet is massive enough to open
a gap in the gas. It happens when the Hill radius of the planet is
larger than 1.5 of the disk scale height. If the mass of the planet is
negligible in comparison to the mass of the disk, the inward velocity of
the planet is regulated by the viscosity of the disk \citep{ward97}:
\begin{equation}
\frac{\mathrm{d} a}{\mathrm{d} t} = - \alpha \frac{C_\mathrm{s}^2}{a \Omega_\mathrm{k}}
\label{e:tIIs}
\end{equation}
where $C_\mr{s}$ is the sound velocity in the gas. In the case of a
planet with a larger mass than the mass of the disk, the migration
is slowed down. The variation in the planetary angular momentum is
then equal to the angular momentum transport rate in the disk \citep{lin96}:
\begin{equation}
\frac{\mathrm{d}}{\mathrm{d} t} [M_\mathrm{p} a^2 \Omega_\mathrm{k}]
= - \frac{3}{2} \nu \Sigma  \Omega_{\rm k} a^2 \ .
\label{e:tIIl}
\end{equation}
 In our calculations we used the smaller of the values of
$\mathrm{d}a/\mathrm{d}t$ given by Eqs. (\ref{e:tIIs}) and
(\ref{e:tIIl}). The migration of the planet was calculated using the
4th order Runge -- Kutta method. It stops when either the time
from the beginning of the calculation (including the time needed for
the formation of the planetesimal swarm) is longer then the lifetime of
the protoplanetary disk $\tau_\mathrm{f}$, or the orbit of the planet becomes
smaller than $0.01~\mr{AU}$. In the latter case we assumed that the planet
is accreted onto the star. In our calculations $\tau_\mathrm{f}=3\times 10^6\
\mr{yrs}$, which  agrees with  observations \citep{haisch}.

\section{Results}
\label{s:results}
Using the methods described in the previous section we were able to
investigate the range of masses and orbits of giant planets which are
able to form in protoplanetary disks characterized by different
initial conditions.  First, we calculated a grid of models of 
protoplanetary disks. Every model is characterized by two parameters:
the initial mass ($M_0$) and outer radius ($R_0$) of the gaseous disk.
 Their initial masses are uniformly distributed
between 0.02 and 0.2 $\mathrm{M_\odot}$. The range of outer radii was
chosen with the goal of including all models in which the formation of giant
planets is possible. The number of radii is constant per interval of
$\lg R_0$.  The viscosity parameter $\alpha$ is equal to 0.001 for all
models. The solids consist of water ice with a bulk density of $1\,
\mathrm{g\,cm^{-3}}$ and have an evaporation temperature of $150\,\mr{K}$. 
For every value of $[M_0, R_0]$ we also checked the gravitational stability
of the corresponding gaseous disk. In some cases the value of the Toomre
parameter
\begin{equation}
  Q=\frac{C_\mathrm{S} \Omega_\mathrm{K}}{\pi G \Sigma}
\end{equation}
drops below 1 in the outer regions of the disk. It means that they are
gravitationally unstable with respect to the axisymmetric modes. We
assumed that these parts of the disk fragment and some giant planets
are formed there on a very short time scale \citep[see
i.e.][]{boss02}. Consequently we used modified initial values of the
mass and outer radius of the disk, which correspond to the mass and
size of the stable part of the original disk. Our distribution the of
initial parameters of protoplanetary disks is an \emph{artificial}
one. In general one should use the distribution that occurs in
nature. Unfortunately current observations do not provide such
information. However, our results remain valid, as we are interested
only in characterizing the general set of possible orbital sizes and
masses of planets.

Using that grid of evolved models we performed Monte-Carlo
calculations to produce the $M_p - a$ distribution of planets
resulting from our models. With the uniform probability we randomly
chose one of the models from the calculated grid. Next, we chose the
initial distance of the planet from the star $a$ with constant
probability per interval of $\lg a$.  We evolved the mass and semimajor
axis of the planet according to the equations given in
Sects. \ref{s:growth} and \ref{s:migr}, starting from the time at which
the planetesimals at radial distance $a$ reach a radius of $2
\mathrm{km}$ in the given model of the protoplanetary disk.  To better
illustrate how the type II migration and the accretion of gas after
opening the gap change the parameters of the giant planets we also
performed calculations in which those two phenomena were not
included.

\begin{figure}
\resizebox{\hsize}{!}{\includegraphics[]{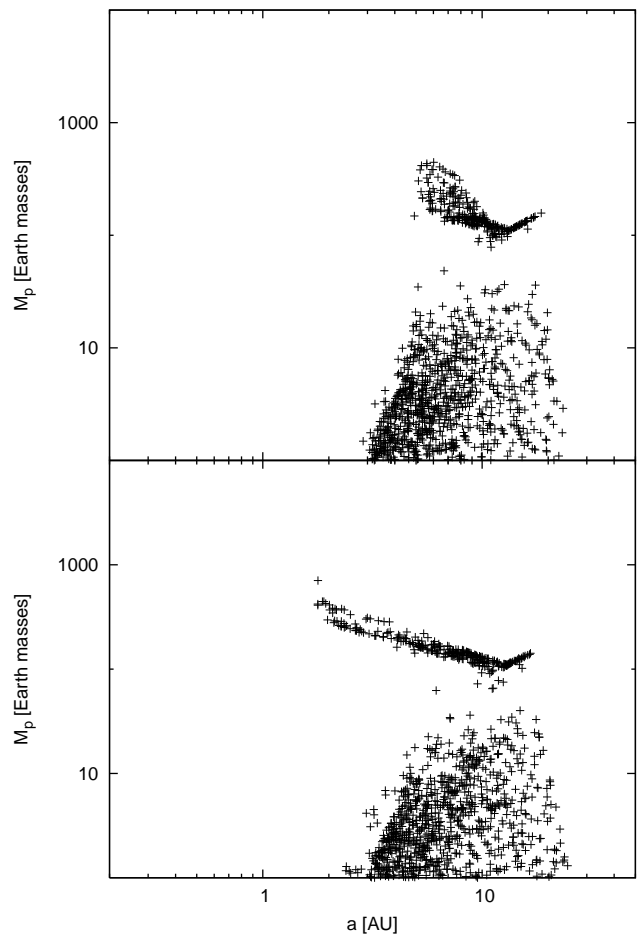}}
\caption{The distribution of theoretically predicted semimajor axes
  and masses of planets around a star with a mass of $1 M_\odot$.  The
  upper panel shows planets that neither migrate nor grow in mass
  after they are massive enough to open the gap in the disk. The lower
  one presents results for planets that undergo type II migration,
  but do not grow after gap opening. The solid material in the disk
  is water ice.}
\label{f:ice}
\end{figure}

The results of calculations in which the planets were neither
permitted to increase their masses after the gap opening  in the
disk or to change their semimajor axes are presented in the upper panel
of Fig. \ref{f:ice}. 
\begin{figure}
\resizebox{\hsize}{!}{\includegraphics[]{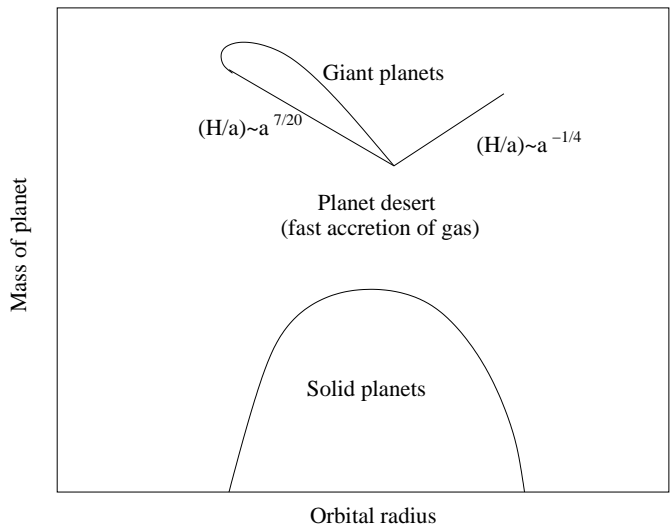}}
\caption{The schematic diagram of  distribution of theoretically predicted semimajor axes
  and masses of planets around a star with a mass of $1 M_\odot$
  without effects of the gap opening and subsequent accretion of gas
  onto the planet.}
\label{f:schem}
\end{figure}
It shows that the planets can be divided into two main groups (see
also Fig. \ref{f:schem}). In the first one there are bodies with
masses smaller than $\sim20 M_\oplus$. These bodies were not able to
accrete a significant amount of gas so the nearly only consist of
their solid cores.  In the second group, there are planets with masses
larger than $\sim100M_\oplus$.  In these cases the runaway gas accretion
 has already taken  place and the majority of their masses are in the form of
gaseous envelopes. They are massive enough to open a gap in the disk.
The small number of planets with intermediate masses results from the
fact that, as the mass of the planet grows above $\sim 20 M_\oplus$, the
timescale of the accretion of gas quickly becomes much shorter than
the lifetime of the protoplanetary disk.  Consequently, the probability of
the disk to disperse before the planet grows sufficiently to open a
gap  is very low \citep[see][]{ida04a}.

The giant planets in this set of models form in a region between 1 and
20~AU from the central star. They can be divided into two subgroups,
each of them forming one of the wings of the V-like shape of the
distribution of giant planets on the $a - M_p$ diagram. As the final
mass of the giant planets in this set of models is determined by
equation
\begin{equation}
M_\mathrm{p}=3 \left(\frac{1.5 H}{a}\right)^{1/3} M_\star \ ,
\end{equation}
 those two subgroups of planets reflect different ways in which the
$H/a$ ratio depends on distance from the star $a$ at the moment of gap
opening. Giant planets at distances larger than $\sim 10 \mr{AU}$ are
at that time in the optically thin parts of the planetary disk in
which $(H/a) \sim a^{7/20}$ \citep[see][]{s98}. Consequently, the
masses of these planets are correlated with their distances from the
star. The correlation is extremely narrow because scale height of the
disk does not depend on its initial mass and outer radius in this
regime in the used model of the gaseous disk.  Giant planets at orbits
smaller than $\sim 10$ AU in the moment of gap opening are in the
optically thick part of the disk in which $(H/a) \sim a^{-1/4}$, so in
their case their masses are anticorrelated with the distance from the
star. The spread of this correlation is larger, because in that case
the exact form of the $H(a)$ relation depends on the initial mass and
radius of the underlying disk. This subgroup contains the most massive,
closest planets formed in this set of simulations. They have masses
around $450 \mathrm{M_\oplus}$ and orbits around 4 AU.

In the next set of models we include the effects of type II migration,
while planets still do not accrete any more after gap opening. The
dependence of the masses of the planets resulting from this
simulations on the sizes of their orbits is presented in the lower
panel of Fig. \ref{f:ice}.  As expected, the only differences in
comparison with the previous case are in the region of the giant
planets. Moreover, the range of the masses of these planets in both
figures are the same. Giant planets at larger orbits open the gap
later because the process of their formation is longer due to lower
densities of planetesimal swarm and smaller Keplerian frequencies (see
Eq. \ref{eq:mc}). Consequently the changes in their orbits are smaller
as they have less time to migrate. Also the accretion rate in the disk
that determines the speed of migration is lower at later times. In
our calculations, only planets at orbits smaller than $\sim 9$ AU are
able to shrink their orbits by a factor larger than 10\%. As the result,
the distribution of planets on the $a - M_p$ diagram for $a> 9$ AU is
nearly the same as in the case without migration. In the case of
planets formed at smaller distances from the star, some of them are
able to migrate to orbits as small as $\sim 2$AU. Additionally, more
massive planets tend to shrink their orbits by a larger factor. This
can be explained by the higher rate of migration in the hotter
disks. At the same time such disks have larger scale heights, and
giant planets can reach larger masses before gap opening and
stopping  their growth. All these factors lead to the small spread
of giant planets around a curve on the $a-M_p$ diagram.
\begin{figure}
\resizebox{\hsize}{!}{\includegraphics{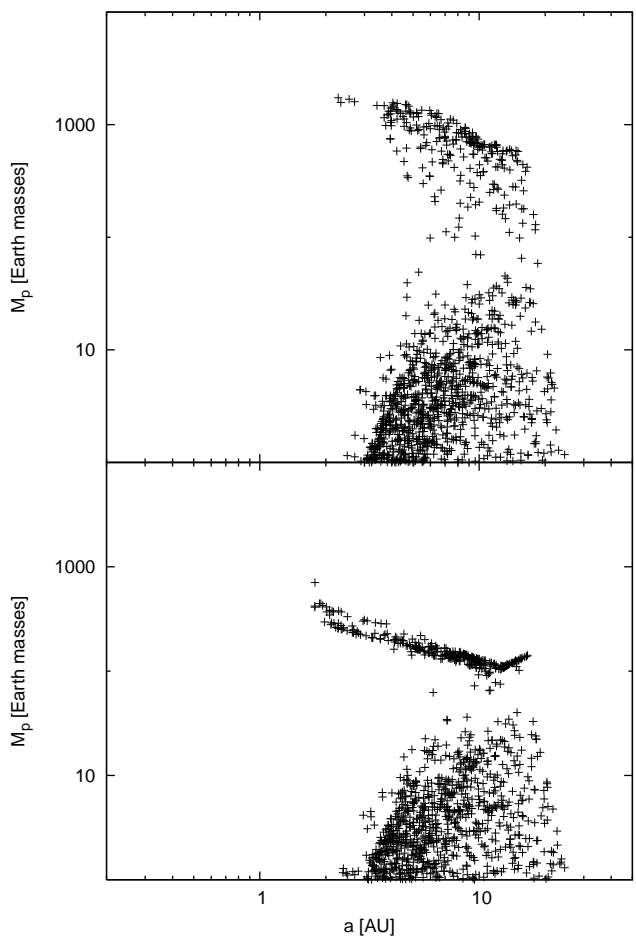}}
\caption{Effect of planet growth after gap opening on the
  theoretically predicted distribution of orbital sizes and masses of
  planets around a star with mass of $1 M_\odot$. The upper panel
  shows the results for planets that undergo type II migration and
  growth after gap opening.  The lower one shows results for planets
  that undergo type II migration, but not grow after gap opening. The
  solid material is water ice.}
\label{f:ice2}
\end{figure}

In the third set of calculations, the planets both migrate and grow in
mass after gap opening.  The distribution of the resulting planets is
shown in Fig. \ref{f:ice2}. The range of their final semimajor axes is
similar to the one in the previous set of models. The maximum size of
planets at the given distance is $\sim 2000 M_\oplus$ at 2 AU and
decreases to $\sim 400 M_\oplus$ at 17 AU. The increase in the masses
of the planets, after gap opening tends to be anticorrelated with
the semimajor axes of their orbits. We explain this fact as follows.
The rate at which a giant planet accretes gas at this stage of its
evolution is determined by the accretion rate in the disk, and as such
does not strongly depend on the distance from the star. At the same
time the protoplanetary cores at smaller distances grow on shorter
timescales. There are two reasons for this. First, the accretion rate
 of planetesimals onto the core is proportional to Keplerian
frequency; second, the surface density of the nascent planetesimal
swarm tends to be larger closer to the star. Consequently,
protoplanets at smaller orbits are able to start accreting
significant amounts of gas for a longer time before dispersion of the
disk. Additionally, the accretion rate in the disk, which determines
the accretion rate of gas onto the planet, is larger in this earlier
epoch. As a result, the maximum size of the planet is everywhere a
decreasing function of the distance from the star, even in those regions
where the relation was the opposite at the moment of gap opening.

\begin{figure}
\resizebox{\hsize}{!}{\includegraphics{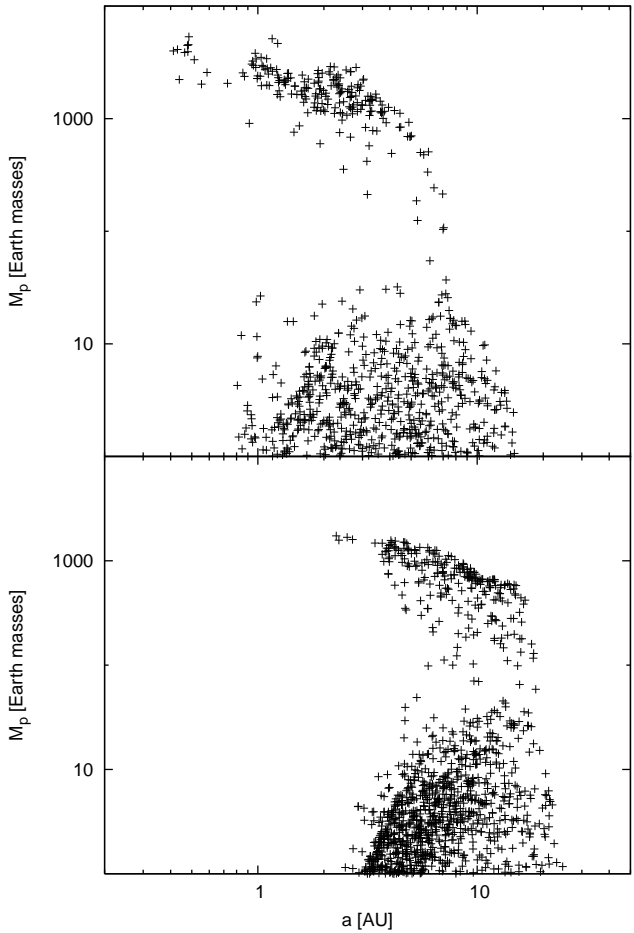}}
\caption{Effect of different kinds of material that the solids in
  the protoplanetary disk consist of on the theoretically predicted
  distribution of orbital sizes and masses of planets around a star
  with a mass of $1 M_\odot$. The upper panel shows the results from models in
  which solids are composed of silicates, the lower one -- from water
  ice.  In both cases the planets undergo type II migration and grow
  after gap opening.  }
\label{f:ht}
\end{figure}

Next, we investigate how the relation between the mass of the planet
and semimajor axes of its orbit depends on the sort of material solids
consist of. For this purpose we performed similar simulations to those
described above, but with solids made of high temperature silicates
instead of water ice. For the evaporation temperature of silicates we
adopted a value of 1350~K, and for their bulk density
$3.3\,\mathrm{g\,cm^{-3}}$. The upper panel of Fig.~\ref{f:ht}
presents the distribution of planets resulting from these
calculations.
Because the solids can survive at higher temperatures than ice, the
giant planets in that case can form closer to the star. Because the
rate of migration does not depend on the composition of solids, 
their final orbits can also be smaller. At the same time, the distribution
of the final surface densities of the planetesimal swarms, as a
function of the distance from the star, seems to be a natural extension
of this distribution in the case of water ice solids. As a result, the
relation of $a - M_p$ for planets with silicate cores at orbits
smaller then 6 AU is just an extension to smaller values of $a$ of the
same relation for planet with water ice cores. 

Consequently, in our models the most massive planets (with $M_p\sim
5000 M_\oplus$) tend to be located closest to the star at orbits with
semimajor axes $\sim ~ 0.5\,\mr{AU}$. On the other hand, planets with
silicate cores at orbits larger than $6\,\mr{AU}$ tend to have smaller
masses than their ice core counterparts. The reason is that at these
distances the time during which the core of the protoplanet only
accretes solid material becomes a significant fraction of the whole
time of its formation. The rate of the accretion of planetesimals onto
the core with the same mass, but composed of more dense material, is
lower due to its smaller physical dimensions. As a result, planets
with silicate cores start their gas accretion later and grow to
smaller masses before the dispersion of the disk. At distances larger
than $\sim 8$ AU, we do not obtain any giant planets with silicate
cores.

\begin{figure}
\resizebox{\hsize}{!}{\includegraphics{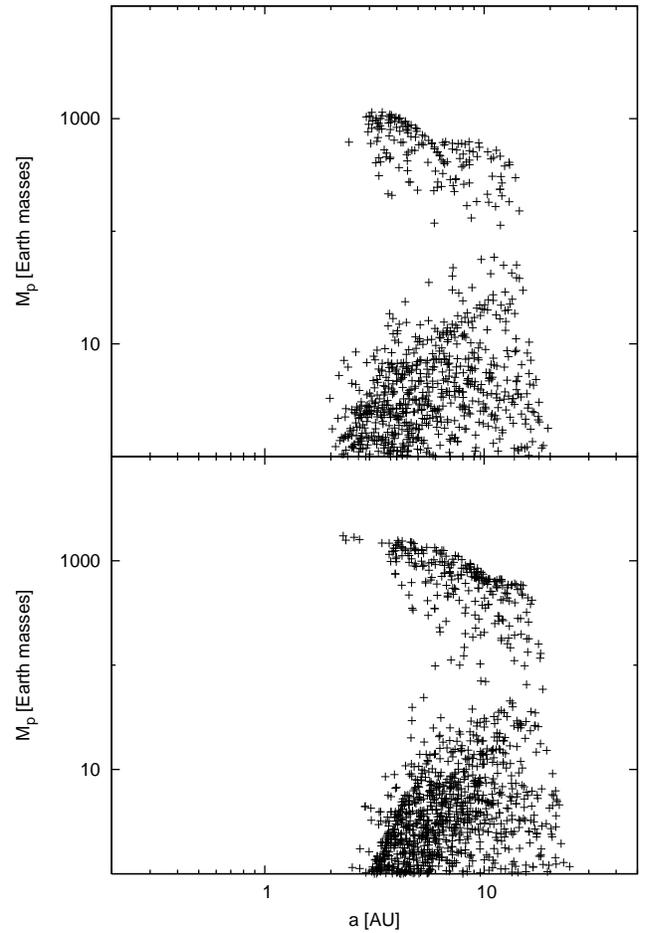}}
\caption{The distribution of theoretically predicted semimajor axes
  and masses of planets around stars with masses of $1 M_\odot$ (lower
  panel) and $0.5 M_\odot$ (upper panel).  In both cases the planets
  undergo type II migration and grow after gap opening.}
\label{f:0.5}
\end{figure}

Finally, we investigated the influence of the central star mass on the
relation between masses and  orbit sizes  of giant planets. In
Fig. \ref{f:0.5} we present the results for two values of the stellar
mass: 0.5 and 1 $M_\odot$.
The planets around less massive stars also tend to be less massive and
have slightly smaller orbits. The first fact is explained by the lower
accretion rates in these disks, as these rates also determine the
maximum rate at which the planet can accumulate gas. The tendency of
giant planets to form around less massive stars at smaller radii is
the result of (a) a more effective radial redistribution of solids in
the disk prior to the formation of planetesimals and, as a result the
higher surface density of the planetesimal swarm, (b) a lower value of
the minimal surface density for the formation of a giant planet around
less massive stars at orbits smaller than $\sim 10$ AU
\citep[see][]{kac_mass}.

\section{Conclusions}
\label{s:concl}
We have presented the results of simulations describing the formation
of giant planets. Our model include all important phases of the
core-accretion scenario, beginning from the formation of the
planetesimal swarm from smaller solids, until the gap opening in the
disk and subsequent planet migration. We presented the distribution of
$a-M_\mr{p}$ for planets resulting from our models for a set of
different initial masses and sizes of the protoplanetary disk and for
the place of planet formation. However, our results cannot be compared
in a simple way to data for extrasolar planets, because of the lack of
knowledge of the distribution of the initial parameters of
protoplanetary disks that occur in nature. Instead our aim was to
characterize the relationship between the final orbital radii and
masses of planets that are in general possible to obtain and to check
how the resulting $a-M_\mr{p}$ distribution changes when including
different physical processes such as migration and gas accretion by
the planet after gap opening. We also investigated whether the
distribution of giant planets is mainly determined by the parameters
of the gaseous disk or the distribution of the planetesimal swarm.

We have shown that, if we do not include the effects of migration nor
accretion of gas by the planet after gap opening, the $a-M_\mr{p}$
distribution for giant planets is mainly determined by the parameters
of the gaseous disk. In that case one is able to read the information
about the dependence of the scale height of the gaseous disk on the
distance from the star from the distribution of giant planets. The
minimal and maximal distance from the star gives information about the
surface density of planetesimal swarm at those places.

By including migration in our models, reconstructing the 
the properties of the gaseous disk from the distribution of giant
planets becomes more difficult. The distance that the  planet can move inward
depends not only on the gaseous environment but also on the 
time between gap opening  and dispersion of the disk. But the
first of these two moments is determined by the surface density of
planetesimal swarm. This, on the other hand, is not related in a simple way
 to the local parameters of the gaseous disk. Our simulations
show that the simple reconstruction of the scale height of the gaseous
disk is possible only from distribution of giant planets at greater distances
 than $\sim 8\,\mr{AU}$, as planets at such large orbits did not
significantly move away from their original places.

If we include in our models  that the planet can still grow in
mass after gap opening, it is even more difficult, if at all possible,
to derive conclusions about the gaseous disk (its temperature, scale
height, etc.) based on the distribution of giant planets.  This is
because the rate of this additional accretion onto the planet depends
mainly on the accretion rate in the disk, while the time frame in
which this accretion can occur is determined by the surface density of
the planetesimal swarm.  The effect of the additional growth of the planet can
even lead  to reversion of the correlation between the masses of the
planets and semimajor axes of their orbits.

We also checked how the  mass of the central star influences the
distribution of planets. The main result is that in disks around less
massive stars, giant planets at the given location tend to be less
massive. At the same time, the giant planets with a given mass tend
to form closer to the less massive stars.

Our models can be seen as an extension of previous calculations
performed by \citet{ida04a} and \citet{alibert05}. In comparison, the
surface density of the planetesimal swarms in our models are computed
self-consistently. This additional factor makes it much more difficult
to predict the  structure of the gaseous disk from the distribution
of giant planets much more difficult. Nevertheless it does seem
necessary. Whenever we performed the similar calculations as
described above, but with the assumption that the surface density of the
planetesimal swarm is always a constant fraction of the surface
density of gas, we were not able to perceive any giant planets. This
seems  contradict  the results of those two papers. A
possible reason for this different result may be the difference in the
underlying models of the gaseous disk. However, we think that the main
factors are the different times for the dispersion of the gaseous disk
(3 Myr vs. 10 Myr) and the different sizes of solids at time 0 (small
grains vs. planetesimal sizes). Our models show that the time needed
to reach planetesimal sizes can be as long 1~Myr and is not
negligible.

Our results also seem to contradic observations that show that more
massive planets tend to be located farther away from the host
star. This correlation is reproduced better by the calculations of
\citet{ida04a}, who conclude that maximal mass of the planets after
including type II migration does not strongly depend on the size of
the orbit. On the other hand, in the results from \citet{alibert05}
the correlation is identical to ours, namely the more massive planets
tend to be closer to the star. This difference can have two
sources. First, \citet{ida04a} neglect the accretion of gas onto the
planet, after it opens a gap in the disk. The second reason, which we
think is more important, can again be the difference in the structure
of the gaseous disk.  In our calculations ratio $(H/r)$ is mainly a
decreasing function of the radius in the inner parts of the disk,
while it is an increasing function in the disk model of
\citet{ida04a}. Because this ratio determines the mass of the planet
which in turn is able to
open a gap in the disk, then it can play a big role in determining the
properties of the set of giant planets.

\begin{acknowledgements}
This project was supported by the German Research Foundation (DFG)
through the Emmy Noether grant WO 857/2-1 and the European
Community's Human Potential Programme trough the contract
HPRN-CT-2002-00308, PLANETS. KK acknowledge the support
from the grant No. 1 P03D 026 26 from the Polish Ministry of
Science.
\end{acknowledgements}

\bibliography{metallicity} 
\bibliographystyle{aa}
\end{document}